\documentclass[preprintnumbers,amsmath,amssymb,nofootinbib,twocolumn]{revtex4}

\usepackage{dcolumn}
\usepackage{bm}

\usepackage{graphicx}

\newcommand{\bea}{\begin{eqnarray}}
\newcommand{\eea}{\end{eqnarray}}

\relax

\begin{document}

\preprint{ANL-HEP-PR-09-16}
\preprint{FERMILAB-PUB-09-097-A}
\preprint{NU-HEP-TH/09-05}

\title{The WIMP Forest: Indirect Detection of a Chiral Square}

\author{Gianfranco Bertone$^a$, C. B. Jackson$^b$, Gabe Shaughnessy$^{b,c}$, Tim M.P. Tait$^{b,c}$, Alberto Vallinotto$^{d}$}
\affiliation{$^a$ Institut d'Astrophysique de Paris, France. UMR7095-CNRS \\ Universit\'e Pierre et Marie Curie, 98bis Boulevard Arago, 75014 Paris, France}
\affiliation{$^b$ Argonne National Laboratory, Argonne, IL 60439}
\affiliation{$^c$ Northwestern University, 2145 Sheridan Road, Evanston, IL 60208}
\affiliation{$^d$ Center for Particle Astrophysics, Fermi National Accelerator Laboratory, 
P.O. Box 500, Kirk Rd. \& Pine St., Batavia, IL 60510-0500 USA}

\begin{abstract}
The spectrum of photons arising from WIMP annihilation carries a detailed
imprint of the structure of the dark sector. In particular,
loop-level annihilations into a photon and another boson can in 
principle lead to a series of lines (a WIMP forest) at energies up to the WIMP 
mass. A specific model which illustrates this feature nicely is a theory of 
two universal extra dimensions compactified 
on a chiral square. Aside from the continuum emission, which is a generic prediction 
of most dark matter candidates, we find a ``forest'' of prominent annihilation lines 
that, after convolution with the angular resolution of current experiments, leads 
to a distinctive (2-bump plus continuum) spectrum, which may be visible in the near 
future with the Fermi Gamma-Ray Space Telescope (formerly known as GLAST).

\end{abstract}

\date{\today}

\maketitle

\section{Introduction}
\label{sec:intro}
Cosmological and astrophysical observations provide compelling evidence for dark matter (DM), 
but so far they have not provided a smoking-gun indication of its identity.
The hunt for dark matter is now entering a new era, with the current 
generation of indirect and direct detection experiments
closing in on weak scale masses and couplings, and the LHC turn-on just around the corner. 
Among the plethora of dark matter candidates, Weakly Interacting Massive Particles (WIMPs)
appear particularly appealing, since they combine the virtues
of weak scale masses and couplings, stability, and connection to the mystery of electroweak
symmetry-breaking ~\cite{reviews}. 

If WIMPs do indeed have weak scale masses and interactions, they are expected to 
produce observable effects in colliders and astroparticle experiments.  In particular,
indirect detection experiments seek secondary particles produced by the annihilation of
WIMPs in regions of our galaxy where the DM density is high, e.g. the Galactic center,
or DM clumps.  Among secondary particles, gamma-rays play a special role, 
since they have the advantage of traveling in straight lines
and without sizable energy losses in the local universe, thus tracing the distribution
of DM in a straightforward way.  

The annihilation of WIMPs into photons typically proceeds via a complicated set of
processes, and the final spectrum actually contains a detailed imprint of
WIMP annihilation that can in principle reveal features such as the WIMP spin and/or other 
particles in the dark sector. Tree-level annihilation of WIMPs into quarks or leptons (or heavier states 
which decay into them) provides a continuum of photon energies, with an upper cutoff 
at approximately the WIMP mass. This continuum emission is in general rather 
featureless, but some models do exhibit spectacular features, such as a high-energy rise 
due to final state radiation \cite{FSR}. 
Furthermore, loop-level annihilation into a photon and $X$ results in a line at energy
\bea
E_{\gamma} = m_{DM} \left( 1 - \frac{M_X^2}{4 m_{DM}^2} \right), 
\eea
where $m_{DM}$ is the WIMP mass and $M_X$ the mass of the other boson in the final 
state (in supersymmetric theories, $X$ is either another photon 
or a $Z$ boson \cite{Bergstrom:1997fh}).  
Line emission typically has smaller magnitude than continuum emission, but the line provides a 
feature that helps discriminate against backgrounds.  More generally, it may also be that the
tree-level annihilation is into photon-unfriendly modes, and the line(s) may turn out to be
prominent features in the photon energy spectrum.

At energies in the 10 GeV -- 10 TeV range, the energy resolution of current gamma-ray
telescopes such as the Fermi LAT or Air Cherenkov Telescopes like CANGAROO, HESS, MAGIC 
and VERITAS, is of order $\Delta E / E \sim 0.1$.
When the mass of $X$ is small compared to the WIMP mass, different lines may merge together because
of finite detector resolutions. 
When $X$ has a mass which is appreciable compared to the WIMP mass,
the lines can be distinct and separately measurable.  When distinct, they provide information
about physics beyond the Standard Model.  For example, the position of the
$\gamma \gamma$ line measures the WIMP mass.  The relative strengths of the $\gamma \gamma$
and $\gamma Z$ lines are a measure of the WIMP coupling to $SU(2)$ singlets versus doublets.
An unsuppressed $\gamma H$ line (where $H$ is a Higgs boson) would be an indication that the 
WIMP is not a Majorana fermion or a scalar.  Finally, if there are particles in the dark sector whose
mass is less than $2m_{DM}$, then a pair of WIMPs may be able to annihilate into one of these exotic heavy
states and a photon.  The existence of such a line could be the first indication that such a heavy
partner is present.  Taken all together, there could be a ``forest" of lines
associated with WIMP annihilations, and they represent a wealth of information about the theory
of dark matter.

A specific model which illustrates these features nicely is the ``Chiral Square" \cite{Dobrescu:2004zi},
a six dimensional model with universal extra dimensions (UED)  \cite{Appelquist:2000nn} -- extra dimensions
in which the entire Standard Model (SM) propagates.  UED theories
contain WIMP candidates because of remnants of the extra-dimensional spacetime symmetries which
forbid the lightest of the Kaluza-Klein excitations of the SM from decaying.  The Chiral Square is an
intrinsically six dimensional construction, whose dark matter candidate, the
``spinless photon"  \cite{Dobrescu:2007ec} 
(the KK mode of the hypercharge vector boson whose spin is pointing in the extra dimension)
has distinct phenomenology compared to the
five dimensional case \cite{Servant:2002aq,Cheng:2002ej,Hooper:2002gs,Bergstrom:2004cy}.  
Six dimensional implementations of the SM
have a mechanism which automatically suppresses dangerous higher dimensional operators leading
to proton decay \cite{Appelquist:2001mj} and motivate the existence of three generations 
of  fermions through gauge anomaly cancellation \cite{Dobrescu:2001ae}.

The article is laid out as follows.  In Section~\ref{sec:ued} we (briefly) review the chiral square model, and 
what is currently known about its dark matter candidate.  
In Section~\ref{sec:continuum}, we calculate the continuum gamma ray spectrum from WIMP 
annihilations in the galaxy. 
Section~\ref{sec:lines} outlines the calculation of the cross sections for the annihilation processes of two WIMPs into states with one or two final primary photons. 
In Section~\ref{sec:prospects} we discuss the prospects for detection of the astrophysical signal and the related uncertainties. We finally reserve Section~\ref{sec:conclusions} for the conclusions.

\section{The Chiral Square}
\label{sec:ued}

The chiral square is a model of two compact universal extra dimensions.  The extra dimensional
coordinates may be represented by a pair of points ($x^5, x^6$) living in a square region 
with sides $L$.  Adjacent sides of the square are identified with each other,
\bea
(y, 0) \equiv (0, y) & ~~~~~  &(y, L) \equiv (L,y)
\eea
which is equivalent to taking a square, folding it along a diagonal, and smoothly gluing the edges together.
The folding leaves the two corners of the square which lie along the fold 
(at $(0,0)$ and $(L,L)$ ) invariant, and identifies the remaining
two corners (at $(0,L)$ and $(L,0)$ ) as the same point.  

The Kaluza Klein (KK) modes of SM fields are labelled by a pair of integers, $(j,k)$ which
satisfy,
\bea
k \geq 0 &, & ~~ j \geq 1-\delta_{k,0} ~.
\eea
Scalar fields have wave functions,
\bea
~~~~~f^{(j,k)} \left( x^5, x^6 \right)  & = &  \frac{1}{L}\left( C_+ + C_- \right), \\
C_{\pm} & = & \cos \pi \left( \frac{j x^5 \pm k x^6}{L} \right),
\nonumber
\eea
with masses given (up to boundary term effects described below) by,
\bea
M^2_{(j,k)} & = & M_0^2 + \pi^2 \frac{j^2 + k^2}{L^2},
\label{eq:KK-masses}
\eea
where $M_0^2$ is the mass of the ``zero-mode" field, which carries no momentum in the 
extra dimensional directions, and has wave function $f^{(0,0)} = 1 / (2 L)$.  We
identify the zero modes with the Standard Model fields.  

The KK modes of fermions are Dirac particles.  The (4D) chirality
associated with the zero mode works essentially as outlined above, but  the opposite chirality pieces have wave
functions which are phase-shifted.  The full details are presented in \cite{Dobrescu:2004zi}, and are not
essential for our purposes.
The gauge bosons $V^M$ decompose into 4D vectors $V^\mu$ and two 4D scalars, $V^5$ and $V^6$.
One linear combination of $V^5$ and $V^6$ is eaten, level by level, by the vector
KK modes to provide their longitudinal degrees of freedom.  
The other linear combination of $V^5$ and $V^6$ are physical gauge 
adjoint scalars in the 4D effective theory.  We follow the usual convention and denote them by
$V_H^{(j,k)}$.

The residual space-time symmetry causes the lightest $(1,0)$ KK particle (LKP) to be stable.
The allowed ``large" interactions can be roughly understood from higher dimensional momentum 
conservation, but with the important observation that since moving across a 
boundary rotates the direction of the momentum by $90^{\circ}$, the momentum in the $x^5$ and $x^6$
directions are not actually distinct from one another. The full momentum conservation, even in this
folded sense, is broken by terms which live on the corners of the square \cite{Ponton:2005kx}
(and whose presence is required when the theory is renormalized \cite{Georgi:2000ks,Cheng:2002iz}),
which nevertheless
preserve a bulk symmetry (invariance under reflections through the middle of the square), provided
the UV physics is such that equal boundary terms live at $(0,0)$ and $(L,L)$.
Even in the presence of such terms, there is a $Z_2$ KK-parity \cite{Dobrescu:2004zi} under which
states that have $j+k$ equal to an even integer are even, and states with $j+k$ odd are odd.
Just as in the 5D case \cite{Cheng:2002iz,Carena:2002me}
the boundary terms have the effect of breaking degeneracies within a given KK level and introducing
KK-number violating (but KK-parity preserving) couplings among the KK modes \cite{Ponton:2005kx}.

We follow the usual reasoning, and assume that the boundary terms are dominated by
radiative contributions from the bulk physics \cite{Cheng:2002iz}.  While not strictly required by
any theoretical or phenomenological argument,\footnote{However, analysis of large
boundary terms in 5D does reveal that when they are large, it is generically
more difficult to fit precision EW data \cite{Flacke:2008ne}.} loop-level boundary terms
motivate having the colored and charged KK modes being heavier than their neutral counterparts
(which, at least at the $(1,0)$ level is essential for the theory to contain a viable WIMP).

Motivated by the expectations of radiative boundary terms, the lightest $(1,0)$ mode is expected to
be the scalar partner of the hypercharge boson, $B_H^{(1,0)}$
(which we will often abbreviate as simply $B_H$) \cite{Ponton:2005kx}.  
While electroweak
symmetry-breaking mixes $B_H^{(1,0)}$ with its $SU(2)$ counterpart, $W_H^{(1,0)}$, the mixing angle
is typically small, and the LKP is, to good approximation, pure $B_H$.  This implies that the LKP
is a gauge singlet, whose coupling to SM fermions and Higgs is controlled by the 
$U(1)$ gauge coupling $g_1$ and the hypercharge of the matter field.  As a further consequence of 
small boundary terms, its largest couplings are expected to be those which are KK number conserving
(in the folded sense above).

The LKP, as a real scalar, has a suppressed annihilation rate to light fermions in the present epoch and therefore annihilates predominantly
into pairs of electroweak bosons $WW$ and $ZZ$, Higgs bosons $HH$, and (if heavy enough) pairs of
top quarks $t \bar{t}$  \cite{Dobrescu:2007ec}.  As a result, its thermal relic density is very sensitive to its mass
and the mass of the zero mode Higgs, which provides a resonant annihilation channel when
$M_H \sim 2 m_{B_H}$.  The relic density is consistent for masses in the range of roughly 200-500~GeV, with
narrow windows of Higgs masses appropriate for each LKP mass.  We use the relic density as a rough
guideline to the most interesting range of LKP masses for indirect detection, but do not strictly assume it applies;
it could be that the LKP is not a thermal relic, or our extrapolation of cosmology to early times is flawed
by imperfect understanding.

\section{Continuum $\gamma$-ray Emission}
\label{sec:continuum}

\begin{figure}[t]
\begin{center}
\includegraphics[width=0.5\textwidth]{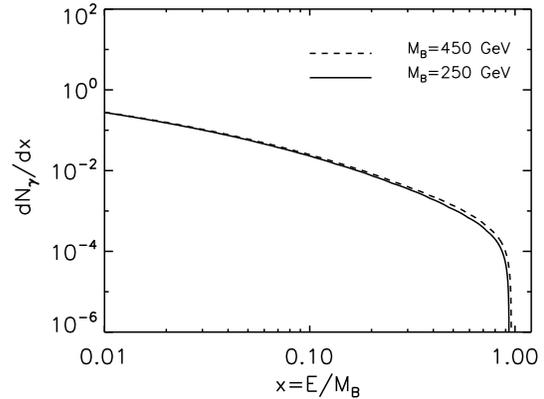} 
\end{center}
\caption[]{Continuum photon spectra, $dN_\gamma/dx$ for two mass choices of $B_H$.}
\label{fg:Continuum}
\end{figure}

As mentioned above, pairs of LKPs annihilate predominantly
into pairs of electroweak bosons $WW$ and $ZZ$, Higgs bosons $HH$, and (if heavy enough) pairs of
top quarks $t \bar{t}$.  Far above the $t \bar{t}$ and $HH$ thresholds, the annihilation
fractions are roughly 
$50\%$ $B_H B_H \rightarrow WW$,
$25\%$ $B_H B_H \rightarrow ZZ$,
$25\%$ $B_H B_H \rightarrow HH$, with
$B_H B_H \rightarrow t \bar{t}$ subdominant.  Continuum gamma ray emission is largely the result
of radiation from charged leptons which result when these massive objects decay, or decays of $\pi^0$s
from the hadronization of strongly interacting decay products.  The result is a rather soft spectrum
of gamma rays, reminiscent of neutralinos in supersymmetric models.

We compute the continuum spectrum using the micrOMEGAs code \cite{Belanger:2007zz}, based on
a CalcHEP \cite{Pukhov:2004ca} (partial) implementation of the chiral square model \cite{kckong}. 
From Fig.~\ref{fg:Continuum}, where the differential flux (per steradian) from the continuum spectrum 
is plotted for $M_{B_H}=250$ GeV and for $M_{B_H}=450$ GeV it is interesting to point out that the 
spectrum sharply decreases well before the value of $M_{B_H}$. This is a distinctive feature of this 
specific chiral square model and is related to the fact that the dominant annihilation channels are
into photon unfriendly modes, consisting of massive (and often neutral) particles which are unlikely
to radiate high energy photons.  This leaves the bulk of the photons coming as radiation (or after
hadronization, decays of $\pi^0$s) from the even softer decay products of the particles produced in the 
primary annihilation, which have less energy available for the final state photons.

\section{Cross Sections for $\gamma$-ray Lines}
\label{sec:lines}

\begin{figure*}[t]
\begin{center}
\includegraphics[bb=95 490 508 735,scale=0.4]{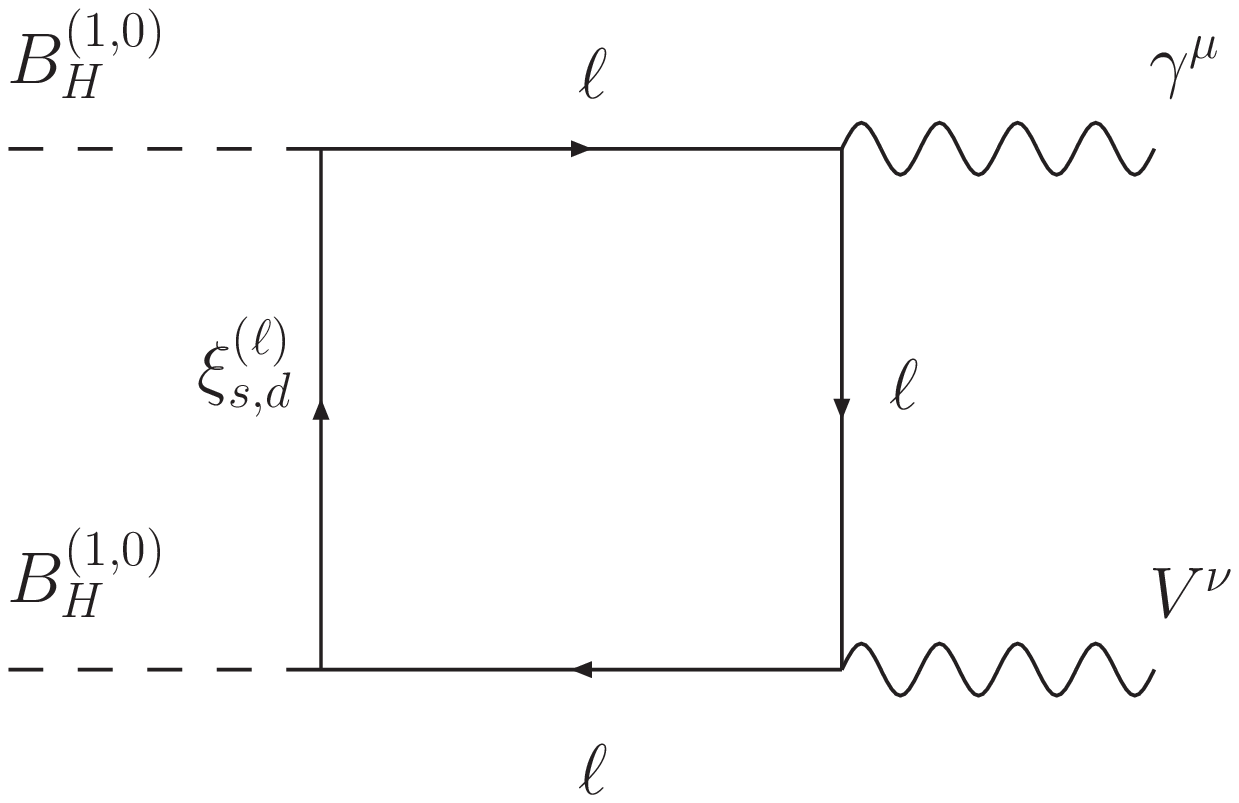} 
\includegraphics[bb=95 490 508 735,scale=0.4]{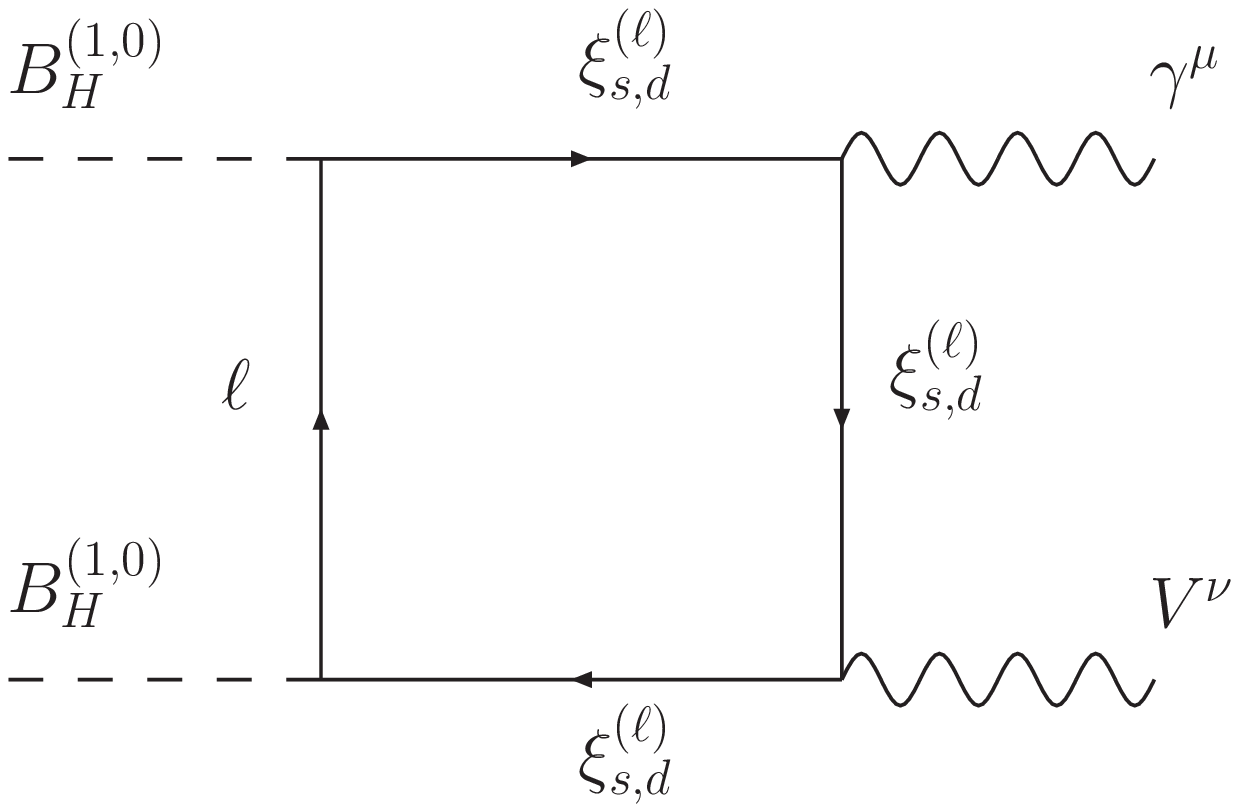} \\
\vspace{0.75cm}
\includegraphics[bb=89 452 503 734,scale=0.4]{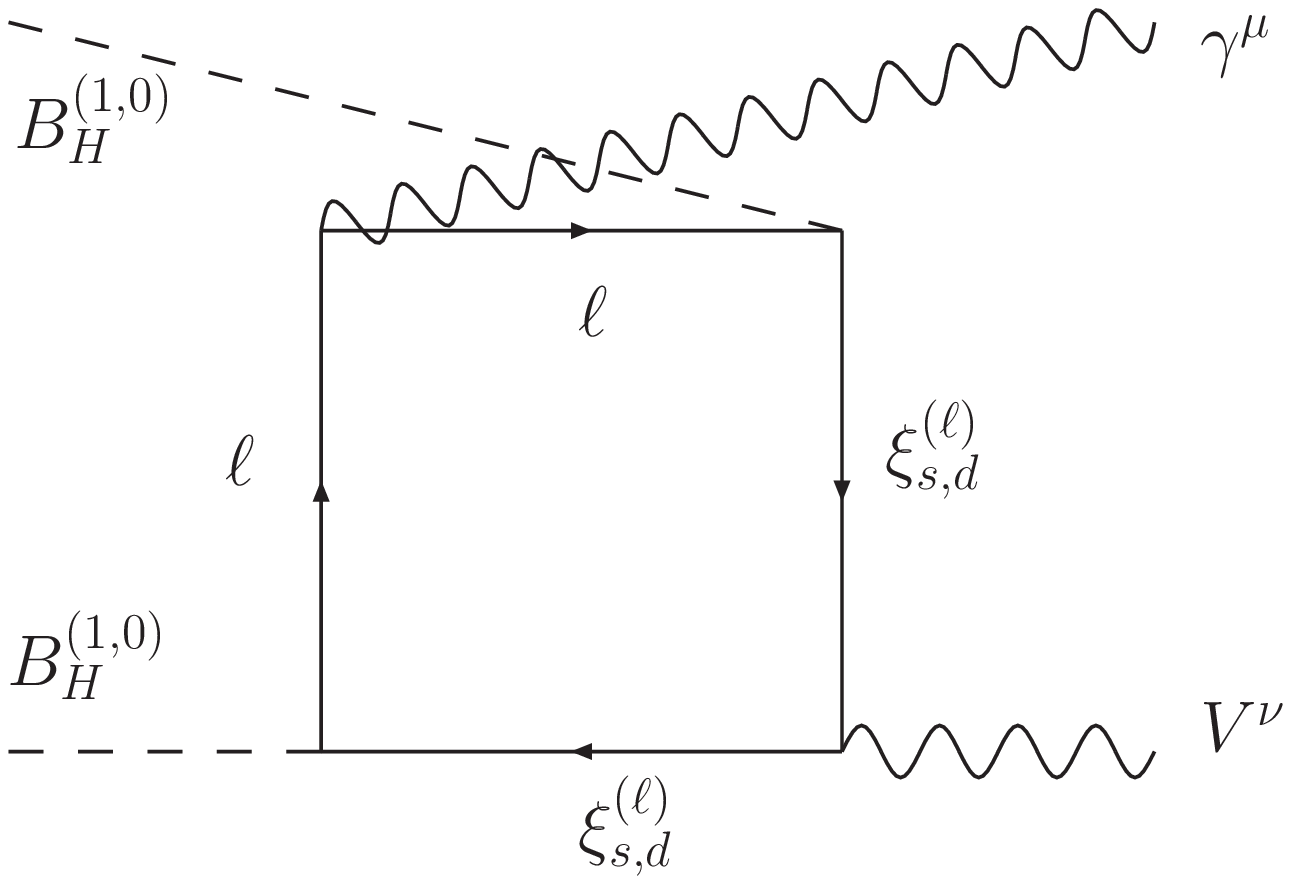}
\includegraphics[bb=89 452 503 734,scale=0.4]{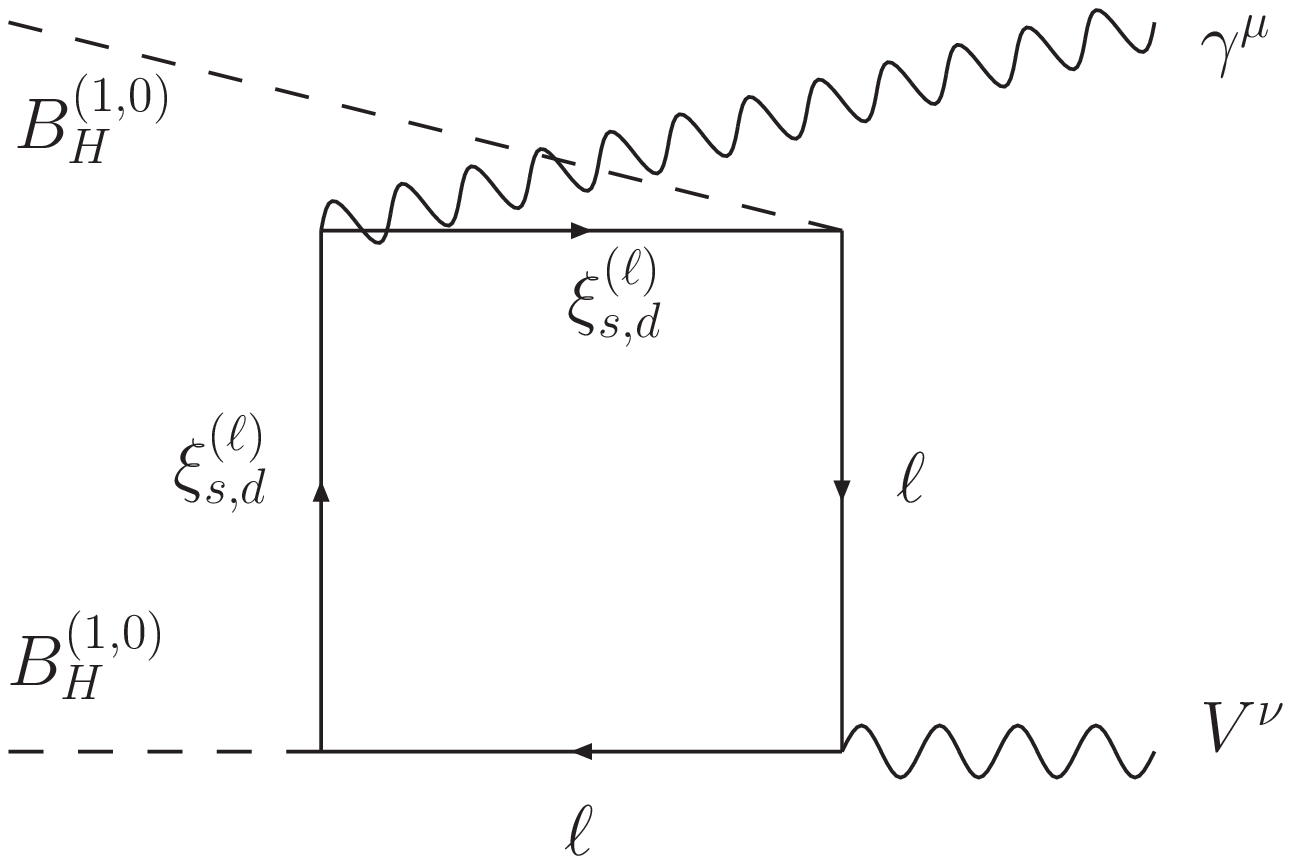}
\end{center}
\caption[]{Examples of Feynman diagrams which contribute to $B_H B_H \to \gamma V$ where $V = \gamma, Z$ and $B^{(1,1)}$.}
\label{fg:feyn-diags}
\end{figure*}

Let us now consider the direct annihilation of LKP pairs into photons.
Specifically, we will consider the process $B_H B_H \to \gamma V$ where $V$
can be either a photon, $Z$ boson or a KK excitation of a SM gauge field.
Since the LKP has no direct coupling to the photon, the leading-order contributions
to these processes occur at one loop.  Examples of Feynman diagrams are shown in 
Fig.~\ref{fg:feyn-diags}.  Here we assume that the dominant
contributions are from loops of SM charged fermions ($\ell$) and their corresponding
pairs of KK partners ($\xi_{s,d}^{(\ell)}$).  Note that, at a given KK level, the
fermionic field content is doubled as compared to the SM such that each SM fermion
has both a singlet ($\xi_{s}^{(\ell)}$) and doublet ($\xi_{d}^{(\ell)}$) KK
partner.

The matrix element for the generic
process $B_H(p_1) B_H(p_2) \to \gamma^\mu(p_A) V^\nu(p_B)$ takes the form:
\begin{equation}
{\cal{M}} = \epsilon_A^{\mu *} (p_A) \epsilon_B^{\nu *} (p_B) {\cal{M}}^{\mu\nu}
 (p_1,p_2,p_A,p_B)\,,
\label{eq:gen-amp}
\end{equation} 
where $\epsilon_A^\mu$ and $\epsilon_B^\nu$ are the polarization tensors of
the photon and $V$ gauge boson, respectively.

In general, the amplitude ${\cal{M}}^{\mu\nu}$ can be expressed as a linear 
combination of tensor structures built from the external momenta and the 
metric tensor $g^{\mu\nu}$.  Considering the transverality of the polarization
tensors, $\epsilon_A \cdot p_A = \epsilon_B \cdot p_B = 0$, the most general
tensor structure is given by:
\begin{eqnarray*}
{\cal{M}}^{\mu\nu} &=& A_1 ~g^{\mu\nu} + B_1 ~p_1^\mu p_1^\nu + B_2 ~p_2^\mu p_2^\nu  
 + B_3 ~p_1^\mu p_2^\nu \nonumber \\ &+& B_4 ~p_1^\nu p_2^\mu +
B_5 ~p_A^\nu p_B^\mu + B_6 ~p_1^\mu p_A^\nu 
  + B_7 ~p_1^\nu p_B^\mu \nonumber \\ &+& B_8 ~p_2^\mu p_A^\nu + B_9 ~p_2^\nu p_B^\mu\,.
\end{eqnarray*}
However, WIMPs are assumed to be highly non-relativistic (with typical 
velocities of $v \sim 10^{-3}$), such that the incoming momenta of the $B_H$'s are 
well approximated by $p_1 = p_2 \equiv p$ where $p = (m_{B_H},\bf{0})$.  Thus, 
using conservation of momentum ($2 p^\mu = p_A^\mu + p_B^\mu$), we can eliminate
many of the above terms.  In addition, since the WIMP's are annihilating nearly 
at rest, the final state products are emitted back-to-back such that (to a 
good approximation):
\begin{equation}
\epsilon_A \cdot p_B = \epsilon_B \cdot p_A = 0.
\end{equation} 
Finally, dot products of the form $\epsilon_{A,B} \cdot p$ will be 
velocity-suppressed and can be safely neglected.  Thus, of the ten original 
tensor structures which make up ${\cal{M}}^{\mu\nu}$, only the $g^{\mu\nu}$
term survives. We have explicitly checked that, indeed, the $A_1$ term is the 
dominant contribution even when all of the tensor structures are kept.

In the calculation of the loop amplitudes, we apply the following algorithm. 
First, we reduce all dot products of the form $k\cdot p$ (where $k$ is the 
loop momentum and $p$ is a generic external momentum) against the corresponding
factors from propagators in the denominator.  The resulting three-point functions can be safely
computed using the standard Passarino-Veltman (PV) technique \cite{Passarino:1978jh}.
The remaining terms in the amplitudes take the form of four-point scalar ($D_0$)
and rank-two tensor ($D_{\mu\nu}$) functions:
\begin{widetext}
\begin{eqnarray}
& & \hspace*{-1.5cm} D_{0 ; \mu\nu}(p_1,p_2,p_3;m_1,m_2,m_3,m_4) =  \nonumber\\
 && \int \frac{d^n k}{i \pi^2}
 \frac{\left\{ 1; k_\mu k_\nu \right\}}{[k^2 - m_1^2] [(k+p_1)^2 - m_2^2] [(k+p_1+p_2)^2 - m_3^2]
 [(k+p_1+p_2+p_3)^2 - m_4^2]} \,,
\label{eq:4pt-integrals}
\end{eqnarray}
\end{widetext}
where $p_i$ are external momenta and $m_i$ are the masses of the particles 
circling the loops.  Note that we have neglected rank-one tensor integrals 
($D_\mu$) since we are only interested in extracting the $g_{\mu\nu}$ pieces
of the amplitude.

Following the PV scheme, the rank-two tensor integral can be rewritten as a linear
expansion in tensor structures which are built from the external momenta and the
metric tensor:
\begin{equation}
D_{\mu\nu} = D_{21} p_{1,\mu}p_{2,\nu}+ D_{22} p_{2,\mu}p_{2,\nu} + \cdots +
D_{27} g_{\mu\nu} \\,.
\end{equation}
The coefficients of this expansion ($D_{ij}$) can then be reduced to scalar 
integrals \cite{Passarino:1978jh}.  However, in cases where two of the external
momenta become identical, as for the case of WIMP annihilation, this approach 
breaks down.  In these cases, the expressions 
for the $D_{ij}$ coefficients in terms of scalar integrals depend inversely on the 
Gram Determinant (GD) built from the external momenta (i.e., 
GD $= det(p_i \cdot p_j)$).  In certain kinematical regions (e.g., where two of 
the momenta become degenerate and GD $\simeq$ 0), the PV scheme gives rise to 
spurious divergences.  In calculations for collider processes (where the momenta 
are integrated over an entire phase space), this situation arises only at 
{\it{special}} points near the boundaries of phase space.  Special techniques involving
interpolating from these {\it{unsafe}} regions of phase space to {\it{safe}} regions
have been developed to deal with these spurious divergences in calculations for
collider processes.  

These techniques do not apply to our situation (where 
the two incoming momenta are fixed {\it{and}} identical) and we are forced to approach 
this problem using the following method.  For our calculation, we have chosen to implement the technique developed in Ref.~\cite{Stuart:1987tt}.  In this algebraic reduction scheme, the original PV scheme
is extended to deal with situations where the GD {\it{exactly}} vanishes.  Higher-point
tensor (and scalar) integrals are expressed in terms of lower-point quantities 
which can be safely evaluated utilizing the usual numerical techniques.  For example,
the expressions for the four-point scalar integral ($D_0$) and the tensor 
coefficient ($D_{27}$) can be expressed as:
\begin{eqnarray}
D_0 &=& \alpha_{123} C_0(123) + \alpha_{124} C_0(124) \nonumber \\ &+& \alpha_{134} C_0(134) +
 \alpha_{234} C_0(234) \,,
\end{eqnarray}
and:
\begin{eqnarray}       
D_{27} &=& \alpha_{123} C_{24}(123) + \alpha_{124} C_{24}(124)  \nonumber \\
 &+& \alpha_{134} C_{24}(134) + \alpha_{234} C_{24}(234) \,,
\end{eqnarray}
where $C_0(ijk)$ and $C_{24}(ijk)$ are the three-point scalar integral and PV
tensor coefficient, respectively (the $(ijk)$ denotes various propagator factors in the 
original four-point denominator).  The $\alpha_{ijk}$ coefficients can
be obtained by solving the matrix equation:
\begin{widetext}
\begin{eqnarray}
\left(
\begin{array}{cccc}
1 & 1 & 1 & 1 \\
0 & p_1^2 & (p_1^2 - p_2^2 + p_5^2)/2 & (p_1^2 + p_4^2 - p_6^2)/2 \\
0 & (-p_1^2 - p_2^2 + p_5^2)/2 & (-p_1^2 + p_2^2 +p_5^2)/2 &
 (-p_1^2 - p_3^2 + p_5^2 + p_6^2)/2 \\
-m_1^2 & p_1^2 - m_2^2 & p_5^2 - m_3^2 & p_4^2 - m_4^2
\end{array}
\right)
\left(
\begin{array}{c}
\alpha_{234} \\
\alpha_{134} \\
\alpha_{124} \\
\alpha_{123} 
\end{array}
\right)
=
\left(
\begin{array}{c}
0 \\
0 \\
0 \\
1 
\end{array}
\right) \,,
\end{eqnarray}
\end{widetext}
where $p_1, \dots, p_4$ are the external momenta, $p_5 = p_1 + p_2$, $p_6 = p_2 + p_3$ and
the $m_i$ are loop particle masses.

This approach allows us to construct quite compact expressions for the 
one-loop amplitudes for the processes of interest 
in terms of kinematical factors and scalar integrals of the form $B_0(p^2;m_1^2,m_2^2)$
and $C_0(p_1^2, p_2^2, p_5^2; m_1^2, m_2^2 , m_3^2)$.  For example, the total amplitude
for $B_H B_H \to \gamma \gamma$ via loops of one (massless) fermion species and its KK partners
is given by:
\begin{widetext}
\begin{eqnarray}
A_1^{(\ell)} &=& - \alpha_Y \alpha_{em} Q_\ell^2 (Y_L^2 + Y_R^2) \biggl\{
 2 + \frac{2}{1 - \eta} B_0(M_{B_H}^2; M_L^2, 0) - B_0(4M_{B_H}^2; 0,0) -
 \frac{1 + \eta}{1 - \eta} B_0(4M_{B_H}^2; M_L^2, M_L^2)  \nonumber\\
 && + 
 M_{B_H}^2 \biggl[ - (1 + \eta)(C_0(M_{B_H}^2,4M_{B_H}^2,M_{B_H}^2;M_L^2,0,0) + 
 C_0(M_{B_H}^2,4M_{B_H}^2,M_{B_H}^2;0,M_L^2,M_L^2)) \nonumber\\
 && - 
 2 C_0(M_{B_H}^2,0,M_{B_H}^2;0,M_L^2,M_L^2) + 
 4\eta C_0(0,0,4M_{B_H}^2;M_L^2,M_L^2,M_L^2) \biggr] \biggr\} \,,
\end{eqnarray}
\end{widetext}
where $\alpha_{em}$ and $\alpha_Y$ denote the SM fine structure constant 
and $U(1)$ coupling constants, respectively.  The charge of the fermion is $Q_\ell$ 
(in units of $e$), while $Y_{L,R}$ are the left- and right-handed hypercharge
quantum numbers of the fermion.  In the above and the following, we assume a common
mass for all KK fermions $M_L$ (i.e., we neglect any mass splittings) and we 
define $\eta \equiv M_L^2/M_{B_H}^2$.  Similar (though more complicated) expressions 
hold for the amplitudes for the processes $B_H B_H \rightarrow \gamma V$ where $V$ is 
a massive gauge boson (e.g., a SM $Z$ boson or a higher KK mode such as the 
$B^{(1,1)}$).

\begin{figure}[t]
\begin{center}
\includegraphics[width=0.5\textwidth]{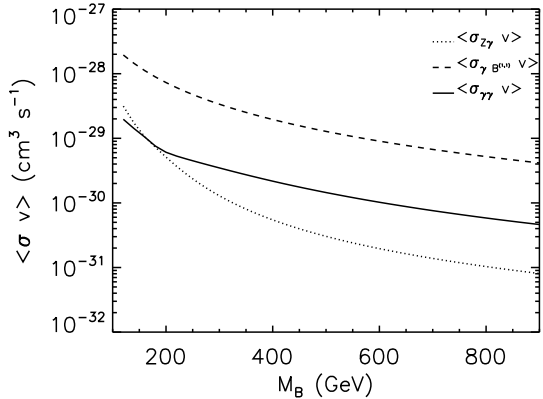} 
\end{center}
\caption[]{Cross sections as a function of the LKP mass $m_{B_H}$ for the three different channels $\gamma\gamma$ (solid), $\gamma Z$ (dotted) and $\gamma B^{(1,1)}$ (dashed).}
\label{fg:xn-vs-mb}
\end{figure}

In Fig.~\ref{fg:xn-vs-mb}, we plot the annihilation cross sections for $\gamma\gamma$, 
$Z\gamma$ and $B^{(1,1)} \gamma$ production as a function of the LKP mass $M_{B_H}$.
Here and in what follows, we have exhanged the compactification scale $L$ for the LKP mass and
then derived the other KK masses ($M_L$ and $M_{B^{(1,1)}}$) in terms of $M_{B_H}$, 
using $M_L= 1.17 M_{B_H}$ and $M_{B^{(1,1)}} = 1.6 M_{B_H}$, but our results are
not strongly dependent on these assumptions unless the mass splittings are unusually large. 
It is interesting 
to note the enhancement of the $B^{(1,1)} \gamma$ cross section compared to the
$\gamma\gamma$ and $Z\gamma$ cross sections.  This effect can be understood as follows.  First, in 
the cases where $V$ is either a photon or a $Z$ boson, the couplings between $V\bar{\ell}\ell$ 
and $V\bar{\xi}\xi$ are nearly the same strength and the result is a significant cancellation
between the various diagrams.  However, in the case where $V$ is identified with the 
$B^{(1,1)}$ KK gauge boson, the $V\bar{\ell}\ell$ couplings are loop-suppressed, while the 
$V\bar{\xi}\xi$
couplings are relatively large.  This results in less cancellation between 
the various diagrams and an enhanced cross section compared to the other two processes.    


\subsection{Line Spectra}

While the spectrum of the $\gamma\gamma$ line is simply a delta function at $M_{B_H}$, the lines arising from the process $B_H \,B_H\rightarrow \gamma V$ will exhibit a line with an intrinsic width, which will depend on the mass of the boson in the final state $M_V$,
\begin{equation}
\frac{dN_{\gamma}^V}{dE}=\frac{4 M_{B_H} M_V \Gamma_V}{f_1 f_2},
\end{equation} 
where $\Gamma_V$ is the $V$ width and
\begin{equation}
f_1 \equiv \left[\tan^{-1}\left(\frac{M_V}{M_{B_H}}\right)
+\tan^{-1}\left(\frac{4M_{B_H}^2-M_V^2}{M_V \Gamma_V}\right)
\right],
\nonumber
\end{equation} 
\begin{equation}
f_2 \equiv \left[\left(4M_{B_H}^2-4M_{B_H}E_{\gamma}-M_V^2\right)^2+\Gamma_V^2M_V^2 \, 
\right].
\nonumber
\end{equation} 
For $B_H\,B_H\rightarrow Z\gamma$, we use the experimentally measured values 
of $M_Z \simeq 91$ GeV and $\Gamma_Z \simeq 2.5$ GeV \cite{Amsler:2008zzb}. 
For $B_H\,B_H\rightarrow B^{(1,1)}\gamma$, we have assumed the minimal mass boundary term
relation $M_{B^{(1,1)}} \approx 1.6 M_{B_H}$.  The width of $B^{(1,1)}$ is determined by its decays into
SM particles.  Such interactions result entirely from boundary terms, and under the minimal
assumption that they are loop-suppressed \cite{Ponton:2005kx}, we expect small coupling to the 
SM such that $\Gamma_{B^{(1,1)}} \approx 10^{-4} M_{B^{(1,1)}}$.  Since $\Gamma_{B^{(1,1)}}$ is typically
much smaller than the typical experimental energy resolutions, the resulting signals at detectors are not
very sensitive to its precise value.

Generally, the larger the mass of $V$, the further away its corresponding line will be from the limiting 
value $M_{B_H}$. The chiral square model is interesting in this respect because the $B^{(1,1)}$ mass 
is large with respect to both $M_{Z}$ and $M_{B_H}$, while still kinematically allowed. This leads to a 
photon distribution for the $B_H\,B_H\rightarrow B^{(1,1)}\gamma$ process that peaks at energies 
that are far below the ones associated with the $Z\gamma$ and $\gamma\gamma$ processes, which
in turn makes the $B^{(1,1)}\gamma$ line clearly distinguishable from the other two lines even by current
experiments, characterised by a relatively large energy resolution.

\section{Prospects for detection}
\label{sec:prospects}

In order to predict the $\gamma$-ray flux from the galactic center generated by the chiral square model, we turn to the evaluation of the spectra for the $B_H \,B_H\rightarrow \gamma V$ processes and of the astrophysical uncertainties related to the integration over the dark matter distribution.
The differential flux of photons arising from dark matter annihilation observed in a direction making an angle $\psi$ with the direction to the galactic center (GC) is given by
\begin{equation}
\frac{d\Phi_{\gamma}}{d\Omega dE}(\psi,E)=\frac{r_{\odot}\rho_{\odot}^2}{4\pi M_{B_H}^2}\frac{dN_{\gamma}}{dE} \int_{\textrm{l.o.s.}}\frac{ds}{r_{\odot}}\,\left[\frac{\rho[r(s,\psi)]}{\rho_{\odot}}\right]^2\label{eq:diff_spectr}
\end{equation} 
with 
\begin{equation}
\frac{dN_{\gamma}}{dE}=\sum_f\langle\sigma v\rangle_f \frac{dN_{\gamma}^f}{dE},
\end{equation} 
where we use the index $f$ to denote the annihilation channels with one or more photons in the final state, $\langle\sigma v\rangle_f$ is the corresponding cross section and $dN_{\gamma}^f/dE$ is the 
(normalized) photon spectrum per annihilation. Furthermore, 
$\rho(\vec{x})$, $\rho_{\odot}=0.3$ GeV/cm$^3$ and $r_{\odot}=8.5$ kpc respectively 
denote the dark matter density at a generic location $\vec{x}$ \textit{with respect to the GC}, 
its value at the solar system location and the distance of the Sun from the GC. Finally, the 
coordinate $s$ runs along the line of sight, which in turn makes an angle $\psi$ with respect 
to the direction of the GC. 

The specification of the dark matter profile is the largest source of uncertainty in the evaluation of the detectability of a dark matter annihilation signal arising from the galactic center, as it fixes the normalization of the predicted flux. The most recent high-resolution numerical simulations show that DM halos can be reasonably well fit with the Navarro Frenk and White (NFW) profile, which is often used as a benchmark for indirect searches 
\cite{Navarro:1995iw}
\begin{equation}
\label{eq:NFW}
\rho_{\rm{NFW}}(r) = \frac{\rho_s}{\frac{r}{r_s}\left( 1 + \frac{r}{r_s}\right)^2} \,.
\end{equation}
However, modifications of the above profile on very small scales 
have been observed in the most recent simulations. In Ref.~\cite{Diemand:2005wv}
it was argued that the innermost regions of DM halos are better 
approximated with $r^{-1.2}$ cusps, while in Ref.~\cite{Navarro:2008kc}
it was found that the analytic form that provides an optimal fit
to the simulated halos is the so-called ``Einasto profile" \cite{Graham:2005xx}
\begin{equation}
\rho(r)=\rho_0\,\exp\left[-\frac{2}{\alpha}\left(\left(\frac{r}{R}\right)^{\alpha}-1\right)\right],
\end{equation} 
which is shallower than NFW at very small radii\footnote{The values assumed for the parameters are in this case $\alpha=0.17$ and $R=20$ kpc.}.

These results have been derived in the framework of simulations containing only dark matter,
not taking baryons into account, which are expected to
play an important role in the dynamics of galaxies, especially on small
scales. Due to dissipative processes, baryons in fact lose energy and 
contract, thus affecting the gravitational potential experienced by 
DM. In the ``adiabatic compression" scenario~\cite{Blumenthal:1985qy}, 
the baryons contraction is quasi-stationary and spherically symmetric.
If one starts from an initial NFW profile, the final slope in the
innermost regions is $r^{-1.5}$~\cite{Edsjo:2004pf,Prada:2004pi,Gnedin:2004cx,Bertone:2005hw}.
Other processes, such as the ``heating" of the DM fluid due to 
gravitational interaction of baryons, may have the opposite effect
on the final DM distribution, thus depleting the central cusps.

Furthermore, the presence of a super-massive black hole (BH) will inevitably 
affect the DM profile, especially within its gravitational radius
(i.e. where the gravitational potential is dominated by the black hole itself). 
The growth of the
BH from an initial small seed would initially lead to a large DM overdensity,
or `spike' \cite{Gondolo:1999ef}. Dynamical effects, and DM 
annihilations, will subsequently tend to destroy the spike \cite{Merritt:2002vj,Ullio:2001fb,Bertone:2005hw}, 
although in some cases significant overdensities can survive over a Hubble time.  
In what follows, we will show our results for two profiles: NFW, and the
adiabatically contracted profile, with the same parameters as the profile 
labelled `A' in Ref.~\cite{Bertone:2005hw}.\footnote{Note that our values
of $J$ for the NFW and ``Adiabatic'' profiles are slightly different from the 
values in Ref. \cite{Bertone:2005hw},
due to the fact that the NFW profile was there approximated as a simple $r^{-1}$
power-law from the galactic center to the location of the Sun.}

\begin{table}[t]
\begin{tabular}{c|c}
\hline
~~~ Model ~~~ & ~~~ $\bar{J}\left( 10^{-5}\right)$ ~~~ \\
\hline\hline
NFW& $1.5 \times 10^4$\\
Adiabatic & $4.7 \times 10^7$ \\
\hline
\end{tabular}
\caption{\label{tab:DM_prof} Value of $\bar{J}(10^{-5})$ for the dark matter density profiles adopted in Fig. \ref{fg:Lines}}
\end{table}


\begin{figure*}[t]
\includegraphics[width=0.49\textwidth]{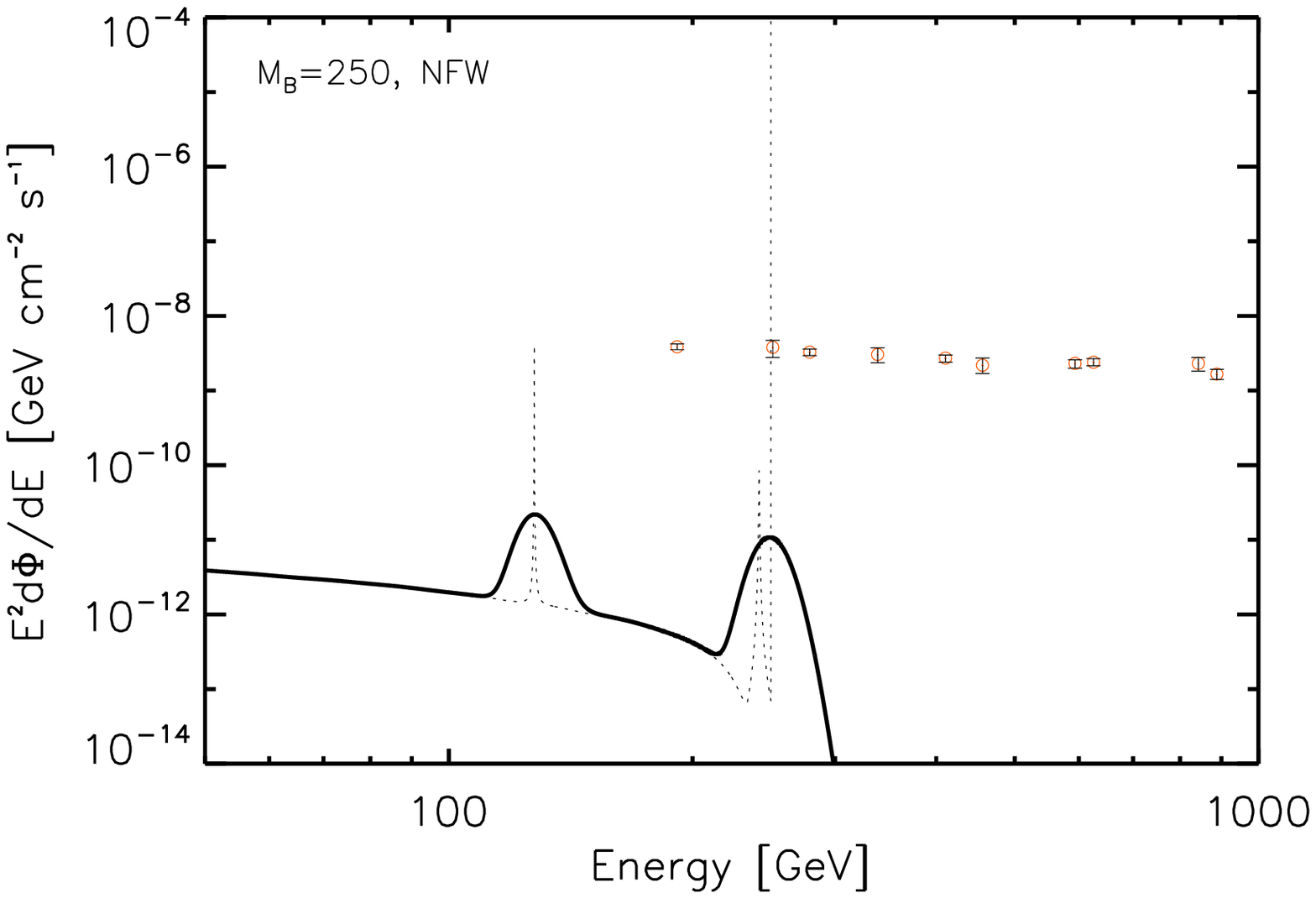}
\includegraphics[width=0.49\textwidth]{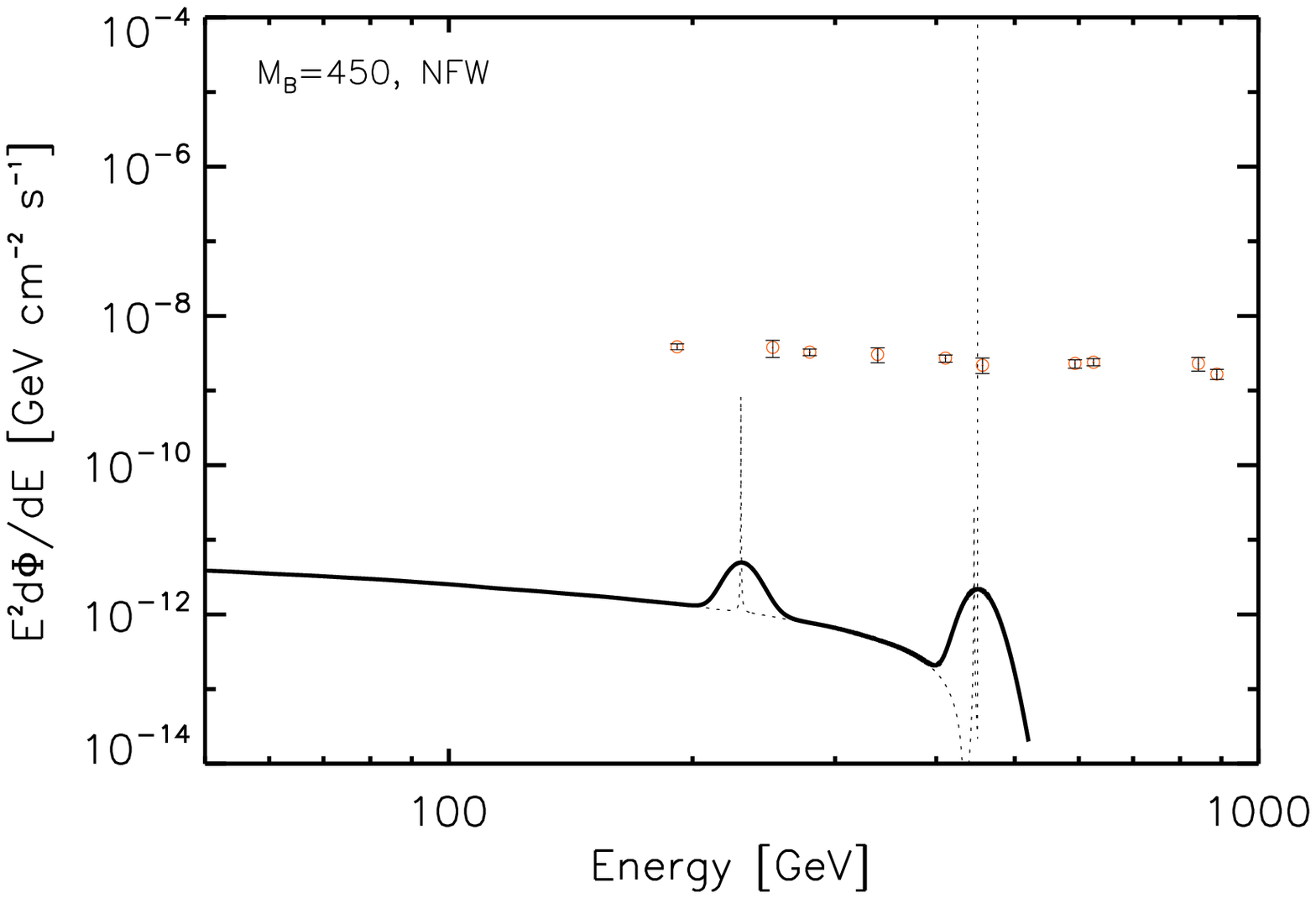}
\includegraphics[width=0.49\textwidth]{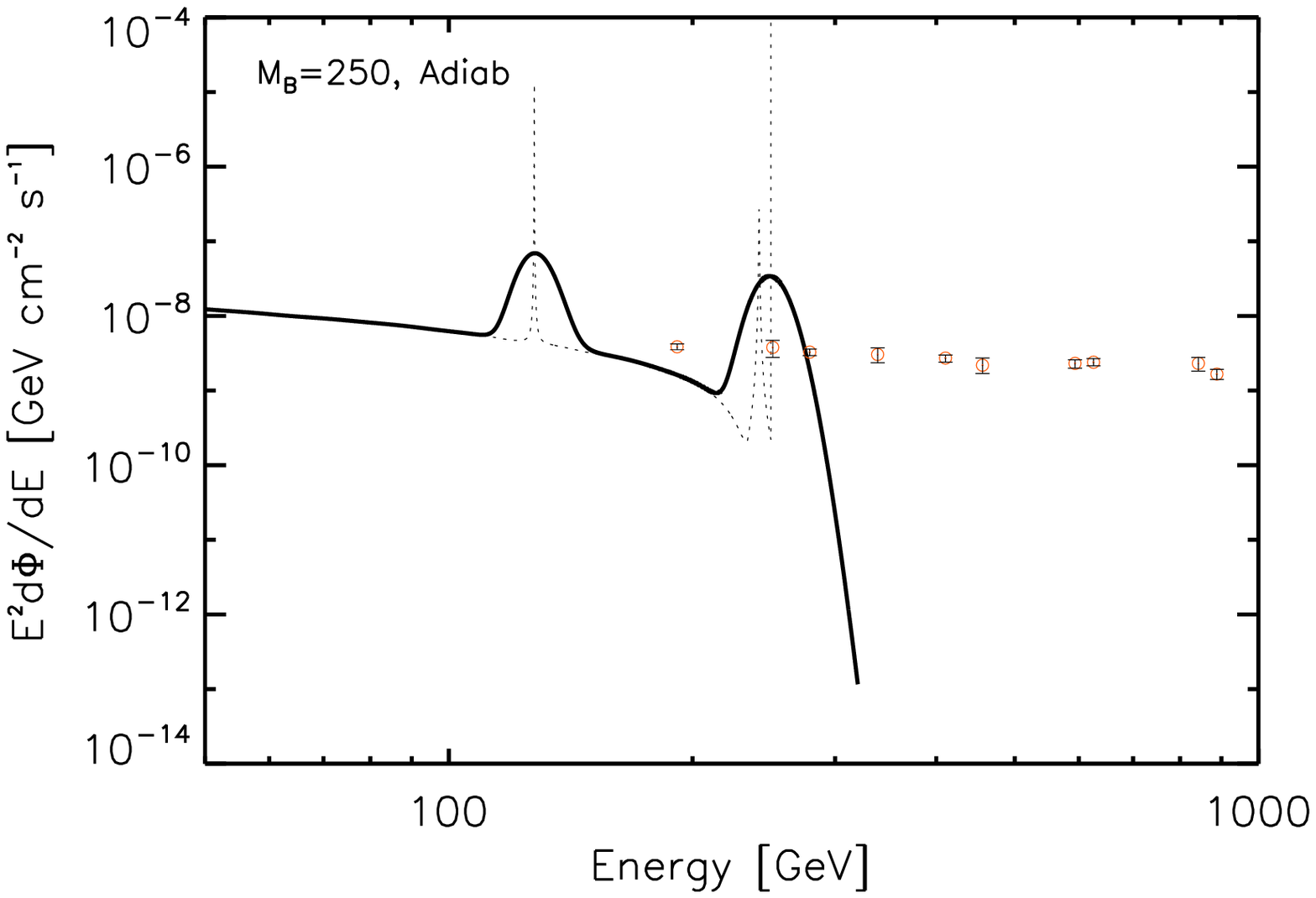}
\includegraphics[width=0.49\textwidth]{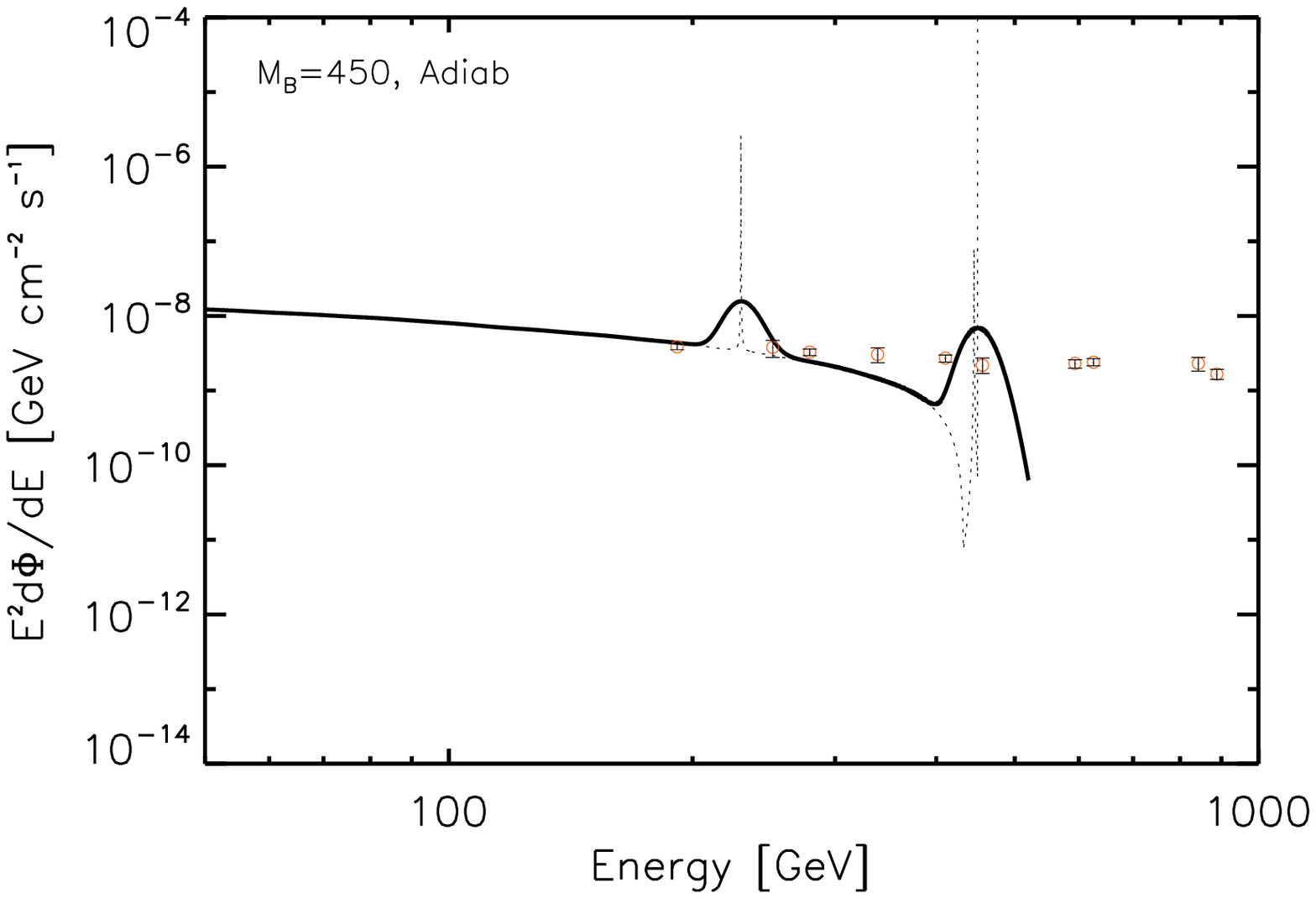}
\caption[]{Predicted fluxes, from a solid angle $\Delta\Omega=10^{-5}$ towards the GC, for the chiral square model with $M_{B_H}=250$ GeV (left column) and $M_{B_H}=450$ GeV (right). We show both the actual spectrum (dotted lines) and the spectrum as it would be observed by an experiment with a 10\% energy resolution (solid) like Fermi LAT. An NFW (adiabatically compressed) profile has been adopted for the lower (upper) panels. We show for reference the HESS data relative to the gamma-ray source detected at the Galactic center.}
\label{fg:Lines}
\end{figure*}

In order to quantify the astrophysical uncertainties related to the dark matter density profile and to separate them from the microphysics, we identify with $J$ the dimensionless integral along the line-of-sight appearing in Eq.~(\ref{eq:diff_spectr}) and with $\bar{J}$ its average value computed for a solid angle $\Delta\Omega$ centered on the GC
\begin{eqnarray}
J&\equiv&\int_{\textrm{l.o.s.}}\frac{ds}{r_{\odot}}\,\left[\frac{\rho[r(s,\psi)]}{\rho_{\odot}}\right]^2,\nonumber\\
\bar{J}(\Delta\Omega)&=&\frac{1}{\Delta\Omega}\int_{\Delta\Omega}J(\psi)\,d\Omega.
\end{eqnarray}
In Tab.~\ref{tab:DM_prof} we show the values of $\bar{J}$ for the two halo profiles adopted in Fig. \ref{fg:Lines}, obtained assuming $\Delta\Omega=10^{-5}$ sr., corresponding to the angular resolution of the HESS and Fermi LAT $\gamma$-ray experiments. 

To account for the finite resolution of the detector we convolve the unfiltered ``raw'' signal $S(E)$ with a gaussian kernel $G(E,E_0)$,
\begin{equation}
G(E,E_0)=\frac{1}{\sqrt{2\pi}E_0\sigma} \exp\left[-\frac{(E-E_0)^2}{2\sigma^2E_0^2}\right],
\end{equation} 
where $\sigma$ is related to the detector's relative energy resolution $\xi$ by $\sigma=\xi/2.3$. The signal $S_M(E_0)$ measured by the detector at energy $E_0$ is then simply given by
\begin{equation}
S_M(E_0)=\int dE\, G(E,E_0)\,S(E).
\end{equation}

\begin{table}[b]
\begin{center}
\begin{tabular}{c|ccc}
\hline
$M_{B_H}$&$L^{-1}$&$M_L$&$M_{B^{(1,1)}}$\\
\hline\hline
  250       &         290      &      294         &         401\\
  450               & 521            &528                 & 721\\
  \hline
\end{tabular}
\end{center}
\caption{\label{tab:masses}Masses of $B_H$, $\xi_{s,d}^{(\ell)}$ and $B^{(1,1)}$ and curvature in GeV for the two mass scenarios given in Fig.~\ref{fg:Lines}.}
\end{table}%
 
We show in Fig.~\ref{fg:Lines} the final predicted fluxes, from a solid angle $\Delta\Omega=10^{-5}$ towards the galactic center, for the chiral square model with $M_{B_H}=250$ GeV and $M_{B_H}=450$ GeV and for two different choices of the DM profile.  Masses of relevant states for these two mass scenarios are given in Tab.~\ref{tab:masses}.  We show both the actual spectrum (dotted lines) and the spectrum as it would be observed by an experiment with a $10\%$ energy resolution (solid) such as Fermi-LAT. We also show for 
comparison the HESS data for the point source HESS J1745-290, which is spatially coincident, within the angular resolution
of HESS, with the supermassive black hole Sgr A* lying at the center of the Galaxy \cite{Aharonian:2006wh}.

\section{Discussion and conclusions}
\label{sec:conclusions}

As can be easily seen from Fig.~\ref{fg:Lines}, the WIMP forest in the chiral square scenario is characterized by three lines. The two lines at and around $M_{B_H}$, belonging respectively to the $B_H\,B_H\rightarrow \gamma\gamma$ and the $B_H\,B_H\rightarrow Z\gamma$ processes cannot be resolved by a detector whose resolution is only 10\% and they appear as a single bump. On the other hand, the $B_H\,B_H\rightarrow B^{(1,1)}\gamma$ process is really the distinctive feature of the chiral square model, leading to a line at much lower energies that is clearly distinguishable from the other ones even by detector with only fair energy resolution. The information content of such a line can be hardly overemphasized. Since 5D Kaluza-Klein models are lacking such a feature, experimental observation of this line would not only make a strong case for Kaluza-Klein dark matter but also indicate the presence of \textit{two} extra dimensions compactified along the lines of the chiral square model and allow a first measurement of the \textit{size} of such extra dimensions.

Figure \ref{fg:Lines} also shows that, for a suitable choice of astrophysical 
and particle physics parameters (especially for an `optimistic' DM profile), the 
annihilation flux in this scenario is within the reach of current 
experiments. The size of 
the error bars in the HESS data, also suggests that the WIMP forest 
(an abbreviated name for the ``forest of gamma-ray lines produced by WIMP annihilations")
might be realistically detected as a ``double bump" in the energy spectrum.
It should be kept in mind, however, that the HESS source actually exhibits 
a flat, feature-less, spectrum extending 
over two decades in energy. This means that the underlying emission 
probably originates from an ordinary astrophysical source, and that this
source represents a foreground for a DM-induced signal. 

A better strategy might consist in the analysis of the gamma-ray emission
from a broader region, e.g. a cone of several degrees towards the 
center, or an annulus that excludes the galactic plane \cite{Stoehr:2003hf,Baltz:2008wd}. A dedicated
analysis of the various detection strategies, also containing a discussion of the
prospects for detecting gamma-ray lines with the Fermi LAT, has been 
recently published in Ref.~\cite{Baltz:2008wd}, where the authors conclude that 
for robust identification, annihilation cross-sections 
to lines as high as $\sigma v \sim 10^{-26}$ cm$^3$ s$^{-1}$ are required,
for an NFW profile. With steeper profiles, even smaller cross sections lead to
annihilation fluxes comparable with, or larger than, astrophysical backgrounds,
thus becoming potentially accessible to gamma-ray telescopes.
In Ref. \cite{Serpico:2008ga}, for instance, it is shown that a suitable choice 
of the observational window may allow to probe cross 
sections of order $\sigma v \sim 10^{-29}$ cm$^3$ s$^{-1}$ for profiles steeper 
than $r^{-1.5}$. These results however strongly depend on the intensity and 
shape of astrophysical foregrounds, that the authors approximated with an 
analytic fit to EGRET data proposed over a decade ago in 
Ref. \cite{Bergstrom:1997fj}. Preliminary Fermi measurements suggest
that EGRET may have overestimated the flux of photons at GeV energies,
therefore a word of caution is in order when estimating the prospects 
for detecting gamma-ray lines in this energy range.

Several other targets for indirect detection have been 
proposed over the years (see Refs. \cite{reviews} for a discussion and 
references). The current understanding is that, aside from a small
region of size $\cal{O}$(1) degrees from the galactic center, the 
diffuse flux is dominated by annihilations in Galactic unresolved 
substructures \cite{Pieri:2007ir,Springel:2008by}, while the extra-galactic 
contribution is expected to be subdominant~\cite{Fornasa:2009qh}.
Alternative targets include clumps of DM (e.g. ~\cite{Pieri:2007ir,Springel:2008by,Baltz:2008wd}
and references therein) and mini-spikes around Intermediate Mass Black Holes
\cite{Bertone:2005xz}.

In closing, gamma rays from the annihilation of dark matter are a fascinating
signal that can tell us a lot about the underlying particle theory, and probe
the distribution of WIMPs in the galaxy.  The lines of the WIMP forest serve
to measure the mass of the WIMP, detect the presence of other states in the
dark sector, and perhaps even identify which theory of dark matter is responsible
for a signal.  While we have worked in the specific context of a six dimensional
UED model, the feature itself is more general, and can be present any time 
two WIMPs can annihilate into a photon and another massive boson.  
As we enter an exciting new era of possibilities to detect WIMPs indirectly, 
it is important to search for multiple features in the spectrum of gamma rays.

\section*{Acknowledgments}

The authors are grateful for conversations with Bogdan Dobrescu, 
Dan Hooper, Rakhi Mahbubani, Simona Murgia, Geraldine Servant, and especially to Eduardo Ponton, 
and K.C. Kong (particularly for his
publicly available chiral square micrOMEGAs model files).
T. Tait is grateful to the SLAC theory group for their extraordinary
generosity during his many visits.
Research at Argonne National Laboratory is
supported in part by the Department of Energy under contract
DE-AC02-06CH11357. A Vallinotto is supported by the DOE and the NASA at Fermilab.


\begin{thebibliography}{99}
\bibitem{reviews}
 G.~Jungman, M.~Kamionkowski and K.~Griest,
 Phys.\ Rept.\  {\bf 267}, 195 (1996)
 [arXiv:hep-ph/9506380];
 L.~Bergstrom,
 Rept.\ Prog.\ Phys.\  {\bf 63}, 793 (2000)
 [arXiv:hep-ph/0002126];
 G.~Bertone, D.~Hooper and J.~Silk,
 Phys.\ Rept.\  {\bf 405} (2005) 279
 [arXiv:hep-ph/0404175].

 \bibitem{FSR}
  L.~Bergstrom, T.~Bringmann, M.~Eriksson and M.~Gustafsson,
  Phys.\ Rev.\ Lett.\  {\bf 94}, 131301 (2005)
  [arXiv:astro-ph/0410359];
  A.~Birkedal, K.~T.~Matchev, M.~Perelstein and A.~Spray,
  arXiv:hep-ph/0507194,
  J.~F.~Beacom, N.~F.~Bell and G.~Bertone,
  Phys.\ Rev.\ Lett.\  {\bf 94} (2005) 171301
  [arXiv:astro-ph/0409403].

\bibitem{Bergstrom:1997fh}
  L.~Bergstrom and P.~Ullio,
  Nucl.\ Phys.\  B {\bf 504}, 27 (1997)
  [arXiv:hep-ph/9706232];
  Z.~Bern, P.~Gondolo and M.~Perelstein,
  Phys.\ Lett.\  B {\bf 411}, 86 (1997)
  [arXiv:hep-ph/9706538];
  P.~Ullio and L.~Bergstrom,
  Phys.\ Rev.\  D {\bf 57}, 1962 (1998)
  [arXiv:hep-ph/9707333];
 L.~Bergstrom, P.~Ullio and J.~H.~Buckley,
 Astropart.\ Phys.\  {\bf 9}, 137 (1998)
 [arXiv:astro-ph/9712318].

\bibitem{Dobrescu:2004zi}
 B.~A.~Dobrescu and E.~Ponton,
 JHEP {\bf 0403}, 071 (2004)
 [arXiv:hep-th/0401032];
 G.~Burdman, B.~A.~Dobrescu and E.~Ponton,
 JHEP {\bf 0602}, 033 (2006)
 [arXiv:hep-ph/0506334];
 G.~Burdman, B.~A.~Dobrescu and E.~Ponton,
 Phys.\ Rev.\  D {\bf 74}, 075008 (2006)
 [arXiv:hep-ph/0601186];
B.~A.~Dobrescu, K.~Kong and R.~Mahbubani,
 JHEP {\bf 0707}, 006 (2007).

\bibitem{Appelquist:2000nn}
 T.~Appelquist, H.~C.~Cheng and B.~A.~Dobrescu,
 Phys.\ Rev.\  D {\bf 64}, 035002 (2001)
 [arXiv:hep-ph/0012100].

\bibitem{Dobrescu:2007ec}
 B.~A.~Dobrescu, D.~Hooper, K.~Kong and R.~Mahbubani,
 JCAP {\bf 0710}, 012 (2007)
 [arXiv:0706.3409 [hep-ph]].

\bibitem{Servant:2002aq}
 G.~Servant and T.~M.~P.~Tait,
 Nucl.\ Phys.\  B {\bf 650}, 391 (2003)
 [arXiv:hep-ph/0206071];
 M.~Kakizaki, S.~Matsumoto, Y.~Sato and M.~Senami,
 Phys.\ Rev.\  D {\bf 71}, 123522 (2005)
 [arXiv:hep-ph/0502059];
 M.~Kakizaki, S.~Matsumoto, Y.~Sato and M.~Senami,
 Nucl.\ Phys.\  B {\bf 735}, 84 (2006)
 [arXiv:hep-ph/0508283];
 F.~Burnell and G.~D.~Kribs,
 Phys.\ Rev.\  D {\bf 73}, 015001 (2006)
 [arXiv:hep-ph/0509118];
 K.~Kong and K.~T.~Matchev,
 JHEP {\bf 0601}, 038 (2006)
 [arXiv:hep-ph/0509119];
 N.~R.~Shah and C.~E.~M.~Wagner,
 Phys.\ Rev.\  D {\bf 74}, 104008 (2006)
 [arXiv:hep-ph/0608140].

\bibitem{Cheng:2002ej}
 H.~C.~Cheng, J.~L.~Feng and K.~T.~Matchev,
 Phys.\ Rev.\ Lett.\  {\bf 89}, 211301 (2002)
 [arXiv:hep-ph/0207125].
 G.~Servant and T.~M.~P.~Tait,
 New J.\ Phys.\  {\bf 4}, 99 (2002)
 [arXiv:hep-ph/0209262].

\bibitem{Hooper:2002gs}
 D.~Hooper and G.~D.~Kribs,
 Phys.\ Rev.\  D {\bf 67}, 055003 (2003)
 [arXiv:hep-ph/0208261];
 G.~Bertone, G.~Servant and G.~Sigl,
 Phys.\ Rev.\  D {\bf 68}, 044008 (2003)
 [arXiv:hep-ph/0211342];
 D.~Hooper and G.~D.~Kribs,
 Phys.\ Rev.\  D {\bf 70}, 115004 (2004)
 [arXiv:hep-ph/0406026];
 T.~Bringmann,
 JCAP {\bf 0508}, 006 (2005)
 [arXiv:astro-ph/0506219];
 A.~Barrau, P.~Salati, G.~Servant, F.~Donato, J.~Grain, D.~Maurin and R.~Taillet,
 Phys.\ Rev.\  D {\bf 72}, 063507 (2005)
 [arXiv:astro-ph/0506389].

\bibitem{Bergstrom:2004cy}
 L.~Bergstrom, T.~Bringmann, M.~Eriksson and M.~Gustafsson,
 Phys.\ Rev.\ Lett.\  {\bf 94}, 131301 (2005)
 [arXiv:astro-ph/0410359].

\bibitem{Appelquist:2001mj}
 T.~Appelquist, B.~A.~Dobrescu, E.~Ponton and H.~U.~Yee,
 Phys.\ Rev.\ Lett.\  {\bf 87}, 181802 (2001)
 [arXiv:hep-ph/0107056].

\bibitem{Dobrescu:2001ae}
 B.~A.~Dobrescu and E.~Poppitz,
 Phys.\ Rev.\ Lett.\  {\bf 87}, 031801 (2001)
 [arXiv:hep-ph/0102010].

\bibitem{Ponton:2005kx}
 E.~Ponton and L.~Wang,
 JHEP {\bf 0611}, 018 (2006)
 [arXiv:hep-ph/0512304].

\bibitem{Georgi:2000ks}
 H.~Georgi, A.~K.~Grant and G.~Hailu,
 Phys.\ Lett.\  B {\bf 506}, 207 (2001)
 [arXiv:hep-ph/0012379].

\bibitem{Cheng:2002iz}
 H.~C.~Cheng, K.~T.~Matchev and M.~Schmaltz,
 Phys.\ Rev.\  D {\bf 66}, 036005 (2002)
 [arXiv:hep-ph/0204342].

\bibitem{Carena:2002me}
 M.~S.~Carena, T.~M.~P.~Tait and C.~E.~M.~Wagner,
 Acta Phys.\ Polon.\  B {\bf 33}, 2355 (2002)
 [arXiv:hep-ph/0207056].

\bibitem{Flacke:2008ne}
 T.~Flacke, A.~Menon and D.~J.~Phalen,
 arXiv:0811.1598 [hep-ph].

\bibitem{Belanger:2007zz}
 G.~Belanger, F.~Boudjema, A.~Pukhov and A.~Semenov,
 Comput.\ Phys.\ Commun.\  {\bf 177}, 894 (2007).

\bibitem{Pukhov:2004ca}
 A.~Pukhov,
 arXiv:hep-ph/0412191.

 \bibitem{kckong}
 {\tt http://home.fnal.gov/~kckong/6d/}

\bibitem{Passarino:1978jh}
 G.~Passarino and M.~J.~G.~Veltman,
 Nucl.\ Phys.\  B {\bf 160}, 151 (1979).

\bibitem{Stuart:1987tt}
 R.~G.~Stuart,
 Comput.\ Phys.\ Commun.\  {\bf 48}, 367 (1988).
 
\bibitem{Amsler:2008zzb}
  C.~Amsler {\it et al.}  [Particle Data Group],
  Phys.\ Lett.\  B {\bf 667}, 1 (2008).

 

\bibitem{Navarro:1995iw}
 J.~F.~Navarro, C.~S.~Frenk and S.~D.~M.~White,
 Astrophys.\ J.\  {\bf 462}, 563 (1996)
 [arXiv:astro-ph/9508025].
 
\bibitem{Diemand:2005wv}
  J.~Diemand, M.~Zemp, B.~Moore, J.~Stadel and M.~Carollo,
  Mon.\ Not.\ Roy.\ Astron.\ Soc.\  {\bf 364}, 665 (2005)
  [arXiv:astro-ph/0504215].
  
\bibitem{Navarro:2008kc}
  J.~F.~Navarro {\it et al.},
  arXiv:0810.1522 [astro-ph].

\bibitem{Graham:2005xx}
 A.~W.~Graham, D.~Merritt, B.~Moore, J.~Diemand and B.~Terzic,
 Astron.\ J.\  {\bf 132}, 2685 (2006)
 [arXiv:astro-ph/0509417].

\bibitem{Blumenthal:1985qy}
 G.~R.~Blumenthal, S.~M.~Faber, R.~Flores and J.~R.~Primack,
 Astrophys.\ J.\  {\bf 301}, 27 (1986).

\bibitem{Edsjo:2004pf}
 J.~Edsjo, M.~Schelke and P.~Ullio,
 JCAP {\bf 0409}, 004 (2004)
 [arXiv:astro-ph/0405414].

\bibitem{Prada:2004pi}
 F.~Prada, A.~Klypin, J.~Flix, M.~Martinez and E.~Simonneau,
 Phys.\ Rev.\ Lett.\  {\bf 93}, 241301 (2004)
 [arXiv:astro-ph/0401512].

\bibitem{Gnedin:2004cx}
 O.~Y.~Gnedin, A.~V.~Kravtsov, A.~A.~Klypin and D.~Nagai,
 Astrophys.\ J.\  {\bf 616}, 16 (2004)
 [arXiv:astro-ph/0406247].
 
\bibitem{Bertone:2005hw}
 G.~Bertone and D.~Merritt,
 Phys.\ Rev.\  D {\bf 72} (2005) 103502
 [arXiv:astro-ph/0501555].

\bibitem{Gondolo:1999ef}
 P.~Gondolo and J.~Silk,
 Phys.\ Rev.\ Lett.\  {\bf 83}, 1719 (1999)
 [arXiv:astro-ph/9906391].

\bibitem{Merritt:2002vj}
 D.~Merritt, M.~Milosavljevic, L.~Verde and R.~Jimenez,
 Phys.\ Rev.\ Lett.\  {\bf 88}, 191301 (2002)
 [arXiv:astro-ph/0201376].

\bibitem{Ullio:2001fb}
 P.~Ullio, H.~Zhao and M.~Kamionkowski,
 Phys.\ Rev.\  D {\bf 64}, 043504 (2001)
 [arXiv:astro-ph/0101481].


\bibitem{Aharonian:2006wh}
 F.~Aharonian {\it et al.}  [H.E.S.S. Collaboration],
 Phys.\ Rev.\ Lett.\  {\bf 97}, 221102 (2006)
 [Erratum-ibid.\  {\bf 97}, 249901 (2006)]
 [arXiv:astro-ph/0610509].

\bibitem{Baltz:2008wd}
 E.~A.~Baltz {\it et al.},
 JCAP {\bf 0807} (2008) 013
 [arXiv:0806.2911 [astro-ph]].

\bibitem{Stoehr:2003hf}
  F.~Stoehr, S.~D.~M.~White, V.~Springel, G.~Tormen and N.~Yoshida,
  Mon.\ Not.\ Roy.\ Astron.\ Soc.\  {\bf 345} (2003) 1313
  [arXiv:astro-ph/0307026].

\bibitem{Serpico:2008ga}
  P.~D.~Serpico and G.~Zaharijas,
  Astropart.\ Phys.\  {\bf 29} (2008) 380
  [arXiv:0802.3245 [astro-ph]].
\bibitem{Bergstrom:1997fj}
  L.~Bergstrom, P.~Ullio and J.~H.~Buckley,
  Astropart.\ Phys.\  {\bf 9} (1998) 137
  [arXiv:astro-ph/9712318].

\bibitem{Pieri:2007ir}
  L.~Pieri, G.~Bertone and E.~Branchini,
  Mon.\ Not.\ Roy.\ Astron.\ Soc.\  {\bf 384} (2008) 1627
  [arXiv:0706.2101 [astro-ph]].
\bibitem{Springel:2008by}
  V.~Springel {\it et al.},
  arXiv:0809.0894 [astro-ph].
\bibitem{Fornasa:2009qh}
  M.~Fornasa, L.~Pieri, G.~Bertone and E.~Branchini,
  arXiv:0901.2921 [astro-ph].


\bibitem{Bertone:2005xz}
  G.~Bertone, A.~R.~Zentner and J.~Silk,
  Phys.\ Rev.\  D {\bf 72} (2005) 103517
  [arXiv:astro-ph/0509565].
 

\end{thebibliography}
\end{document}